\documentclass[
	a4paper, 
	10pt, 
	unnumberedsections, 
	twoside, 
]{LTJournalArticle}

\usepackage[acronym]{glossaries}
\usepackage{siunitx}
\usepackage{tikz}
\usepackage{lettrine}

\newcommand{\circlednum}[1]{%
  \tikz[baseline=(char.base)]{
    \node[shape=circle, draw, inner sep=1pt] (char) {#1};
  }%
}

\runninghead{} 

\footertext{\textit{RISC-V Summit Europe, Paris, 12-15th May 2025}} 

\title{Fused-Tiled Layers: Minimizing Data Movement on RISC-V SoCs with Software-Managed Caches} 

\author{%
	Victor J.B. Jung\textsuperscript{$\dagger$}, Alessio Burrello\textsuperscript{$\ddagger$}, Francesco Conti\textsuperscript{*} and Luca Benini\textsuperscript{$\dagger$}\thanks{This work has received funding from the Swiss State Secretariat for Education, Research, and Innovation (SERI) under the SwissChips initiative.}
}

\date{\footnotesize\textsuperscript{\textbf{$\dagger$}}Integrated Systems
Laboratory (IIS), ETH Zürich\\ \textsuperscript{\textbf{*}}Department of Electrical,
Electronic and Information Engineering, University of Bologna\\ \textsuperscript{\textbf{$\ddagger$}} Interuniversity Department of Regional and
Urban Studies and Planning, Politecnico di Torino}

\newacronym[plural=DNNs, firstplural=Deep Neural Networks (DNNs)]{dnn}{DNN}{Deep Neural Network}
\newacronym[plural=SoCs, firstplural=Systems-on-chip (SoCs)]{soc}{SoC}{System-on-chip}
\newacronym[plural=ONNXs, firstplural=Open Neural Network Exchanges (ONNXs)]{onnx}{ONNX}{Open Neural Network Exchange}
\newacronym[plural=TCFs, firstplural=Tile Constraint Flows (TCFs)]{tcf}{TCF}{Tile Constraint Flow}
\newacronym[plural=TCs, firstplural=Tile Constraints (TCs)]{tc}{TC}{Tile Constraint}
\newacronym[plural=GEMMs, firstplural=General Matrix Multiplications (GEMMs)]{gemm}{GEMM}{General Matrix Multiplication}
\newacronym{cp}{CP}{Constraint Program}
\newacronym[plural=QNNs, firstplural=Quantized Neural Networks (QNNs)]{qnn}{QNN}{Quantized Neural Network}
\newacronym[plural=FMs, firstplural=Foundation Models (FMs)]{fm}{FM}{Foundation Model}
\newacronym{xr}{XR}{Extended Reality}
\newacronym[plural=RNNs, firstplural=Recurrent Neural Networks (RNNs)]{rnn}{RNN}{Recurrent Neural Network}
\newacronym[plural=CNNs, firstplural=Convolutional Neural Networks (CNNs)]{cnn}{CNN}{Convolutional Neural Network}
\newacronym{nlp}{NLP}{Natural Language Processing}
\newacronym{hpc}{HPC}{High-Performance Computing}
\newacronym{eeg}{EEG}{Electroencephalography}
\newacronym[plural=LLMs, firstplural=Large Language Models (LLMs)]{llm}{LLM}{Large Language Model}
\newacronym[plural=SLMs, firstplural=Small Language Models (SLMs)]{slm}{SLM}{Small Language Model}
\newacronym[plural=GPGPUs, firstplural=General-Purpose Graphics Processing Units (GPGPUs)]{gpgpu}{GPGPU}{General-Purpose Graphics Processing Unit}
\newacronym[plural=TPUs, firstplural=Tensor Processing Units (TPUs)]{tpu}{TPU}{Tensor Processing Unit}
\newacronym[plural=DMAs, firstplural=Direct Memory Access (DMAs)]{dma}{DMA}{Direct Memory Access}
\newacronym{ai}{AI}{Artificial Intelligence}
\newacronym[plural=MLs, firstplural=Machine Learnings (MLs)]{ml}{ML}{Machine Learning}
\newacronym[plural=MMUs, firstplural=Memory-Management Units (MMUs)]{mmu}{MMU}{Memory-Management Unit}
\newacronym[plural=ISAs, firstplural=Instruction Set Architectures (ISAs)]{isa}{ISA}{Instruction Set Architecture}
\newacronym{simd}{SIMD}{Single Instruction Multiple Data}
\newacronym{spmd}{SPMD}{Single Program Multiple Data}
\newacronym{dsp}{DSP}{Digital Signal Processing}
\newacronym[plural=MCUs, firstplural=Microcontrollers (MCUs)]{mcu}{MCU}{Microcontroller}
\newacronym{tcdm}{TCDM}{Tightly-coupled Data Memory}
\newacronym{nms}{NMS}{Neural Memory Subsystem}
\newacronym{npu}{NPU}{Neural Processing Unit}
\newacronym{mram}{MRAM}{Magnetoresistive Random Access Memory}
\newacronym{sram}{SRAM}{Static Random Access Memory}
\newacronym{axi}{AXI}{Advanced eXtensible Interface Bus}
\newacronym{ptq}{PTQ}{Post-Training Quantization}
\newacronym{qat}{QAT}{Quantization-Aware Training}
\newacronym{efm}{EFM}{Embodied Foundation Model}
\newacronym{ast}{AST}{Abstract Syntax Tree}
\newacronym{tqt}{TQT}{Trained Quantization Thresholds}
\newacronym{api}{API}{Application Programming Interface}
\newacronym{cim}{CIM}{Compute-In Memory}
\newacronym{mps}{MPS}{Metal Performance Shaders}
\newacronym[plural=MPUs, firstplural=Microprocessors (MPUs)]{mpu}{MPU}{Microprocessor}
\newacronym{dram}{DRAM}{Dynamic Random Access Memory}
\newacronym{ann}{ANN}{Artifical Neural Network}
\newacronym{fc}{FC}{Fully-Connected}
\newacronym{vit}{ViT}{Vision Transformer}
\newacronym{ftl}{FTL}{Fused-Tiled Layers}
\newacronym{mlp}{MLP}{Multi-Layer Perceptron}
\newacronym{ram}{RAM}{Random-Access Memory}
\newacronym{llc}{LLC}{Last Level Cache}

\renewcommand{\maketitlehookd}{%
	\begin{abstract}
		\noindent The success of \glspl{dnn} and their high computational requirements pushed for large codesign efforts aiming at \gls{dnn} acceleration. 
        Since \glspl{dnn} can be represented as static computational graphs, static memory allocation and tiling are two crucial optimizations. Hence, \glspl{soc} specialized for \gls{dnn} acceleration commonly features a multi-level software-managed memory hierarchy. 
        In such architecture, layer-wise tiling, i.e., splitting each layer into multiple sub-nodes, is commonly used; however, while reducing memory occupation, it can increase the total memory transfer, ultimately causing costly off-chip memory copies, which impact energy efficiency and create memory bottlenecks.
        This work proposes \gls{ftl}, a novel algorithm for automatic fusion between tiled layers.
        We leverage the flexibility and efficiency of a RISC-V (RV32) heterogeneous \gls{soc} to integrate \gls{ftl} in an open-source deployment framework, which we tune for RISC-V targets. 
        We demonstrate that \gls{ftl} brings up to 60.1\% runtime reduction for a typical \gls{mlp} stage of \glspl{vit} due to the reduction of off-chip transfer and on-chip data movement by 47.1\,\%.
    \end{abstract}
}

\begin{document}

\maketitle 

\section{Introduction}
\vspace{-1em}

\lettrine[]{W}{ith} the increased interest in efficient execution of \glspl{dnn} at the edge, hardware architectures and \gls{dnn} compilers have been proposed. 
While ARM-based architectures have a high level of maturity in both hardware and software ecosystems that allows for optimized execution of modern \gls{dnn} models, the RISC-V ecosystem has proliferated in terms of novel hardware architectures, but there is room for improvement in terms of software support.
%

%
Specifically, while popular \gls{dnn} compilers such as CubeAI\footnote{\url{https://stm32ai.st.com/stm32-cube-ai/}}, TVM~\cite{chen2018tvm}, or IREE\footnote{\url{https://github.com/iree-org/iree}} present streamlined support for ARM-based \glspl{soc}, important features of modern RISC-V \glspl{soc}, such as multi-level memory hierarchy, are not considered.
In particular, layer fusion is a well-known technique performed by \gls{dnn} compilers to avoid materializing intermediate tensors that increase memory footprint and bandwidth.
In RISC-V-based platforms with multi-level memory hierarchy, it is particularly critical to avoid materializing huge intermediate activations in \gls{llc} memory.
Therefore, this paper presents a new flexible algorithm tailored to RISC-V \glspl{soc} with a multi-level memory hierarchy, \textit{Fused-Tiled Layers (FTL)}, to minimize memory transactions by fusing a series of layers:
\begin{enumerate}

\item \gls{ftl} formulates the tiling of each \gls{dnn} layer as a constraint optimization problem, where each output tensor dimension is linked to input tensor dimensions via a linear transformation, allowing us to merge several layers to generate valid layer fusion solutions for any layer combination. By doing so, we minimize transfers from L2 memory to \gls{llc}.

\item We benchmark \gls{ftl} with a \gls{vit}'s \gls{mlp} on a reduced version of the heterogeneous RISC-V \gls{soc} Siracusa~\cite{prasad2024siracusa}, with, and without the \gls{npu}. Compared to the layer-per-layer tiling strategy, we demonstrate a runtime reduction of 28.8\,\% when only using the 8-cores cluster of RISC-V cores and 60.1\,\% when using the cluster and the \gls{npu}.
\end{enumerate}

\begin{figure*}[hbt!]
  \includegraphics[width=\linewidth]{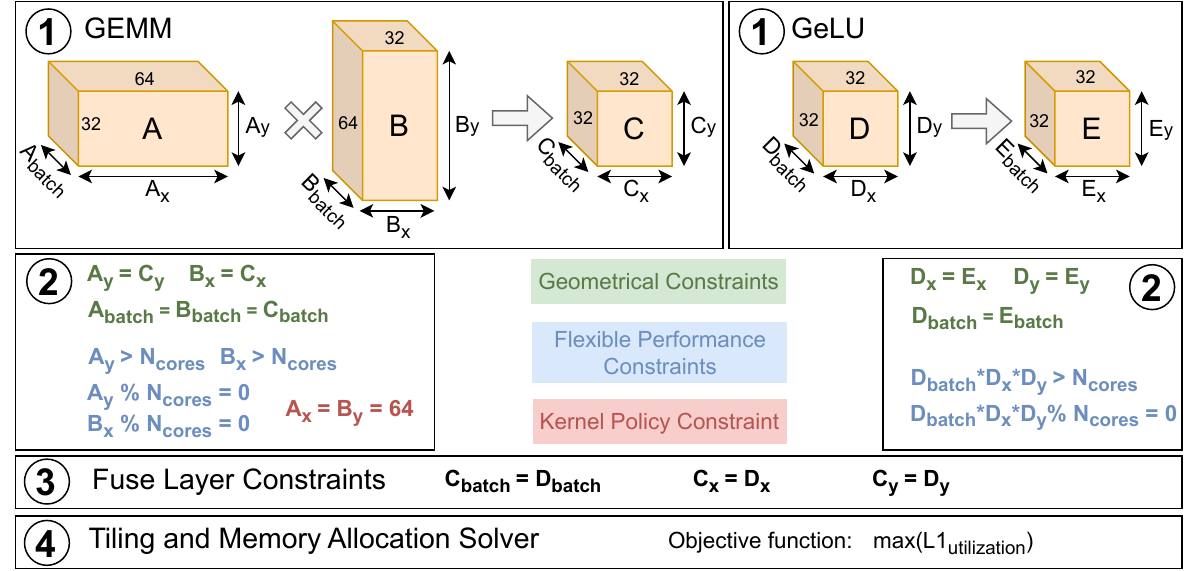}
  \caption{Overview of the \gls{ftl} on a \gls{gemm} and GeLU layer.}
  \label{fig:FTL-overview}
  \vspace{-1em}
\end{figure*}

\vspace{-2em}
\section{Methodology}
\vspace{-1em}

We integrate \gls{ftl} into Deeploy\footnote{\url{https://github.com/pulp-platform/Deeploy}}, an open-source bottom-up \gls{dnn} deployment framework that generates optimized bare-metal C requiring minimal runtime support. 
We rely on kernels using the extended \textit{RV32IMCF-XpulpV2} \gls{isa} featuring hardware loops, post-increment load-store, and \gls{simd} instructions.
To move tiles of tensors across the memory hierarchy, we use \gls{dma} engines that rely on the flexibility of RISC-V systems to perform 3D transfers. 
Fig~\ref{fig:FTL-overview} provides a visual representation of the different steps (numbered from 1 to 4) of \gls{ftl}. 
%
In step \circlednum{1}, we attribute a variable for each tensor dimension related to the given operator. 
Then, we formulate the constraints for the tiling of the single operator in step \circlednum{2}. There are three kinds of constraints: the \textit{geometrical constraints} describe the data dependency between the dimensions of the output and input tensors. The \textit{kernel policy constraints} are specific to the kernel's dataflow; 
Finally, we add \textit{flexible performance constraints} to boost the hardware utilization.
In step \circlednum{3}, we select the consecutive layers to fuse and bind the variable of their shared tensors dimension. 
Finally, in step \circlednum{4}, we solve the constraint optimization problem representing the fused layers with Deeploy's tiling and memory allocation solver.

\begin{figure}[t]
  \centering
  \includegraphics[width=\linewidth]{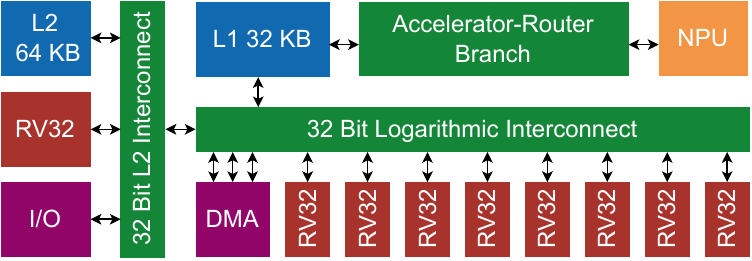}
  \caption{Overview of the modified Siracusa \gls{soc}.}
  \label{fig:siracusa}
  \vspace{-1.5em}
\end{figure}

\vspace{-1em}
\section{Results}
\vspace{-1em}

\subsection{Evaluation Setup}

We perform our benchmark on a reduced version of the RISC-V Siracusa~\cite{prasad2024siracusa} \gls{soc}; its architecture is described in Fig~\ref{fig:siracusa}.
%
The 8 RISC-V cores are using the \textit{RV32IMCF-XPulpV2} \gls{isa} tailored to \gls{dsp} tasks, and the \gls{npu} is targeting \gls{gemm} and convolution. 
We use the GVSoC\footnote{\url{https://github.com/gvsoc}} event-based simulator to measure the runtime, which provides fast and accurate simulation with an error typically below 10\,\%.

\vspace{-1em}
\subsection{\gls{vit}'s \gls{mlp} benchmark}

\begin{figure}[t]
  \centering
  \includegraphics[width=\linewidth]{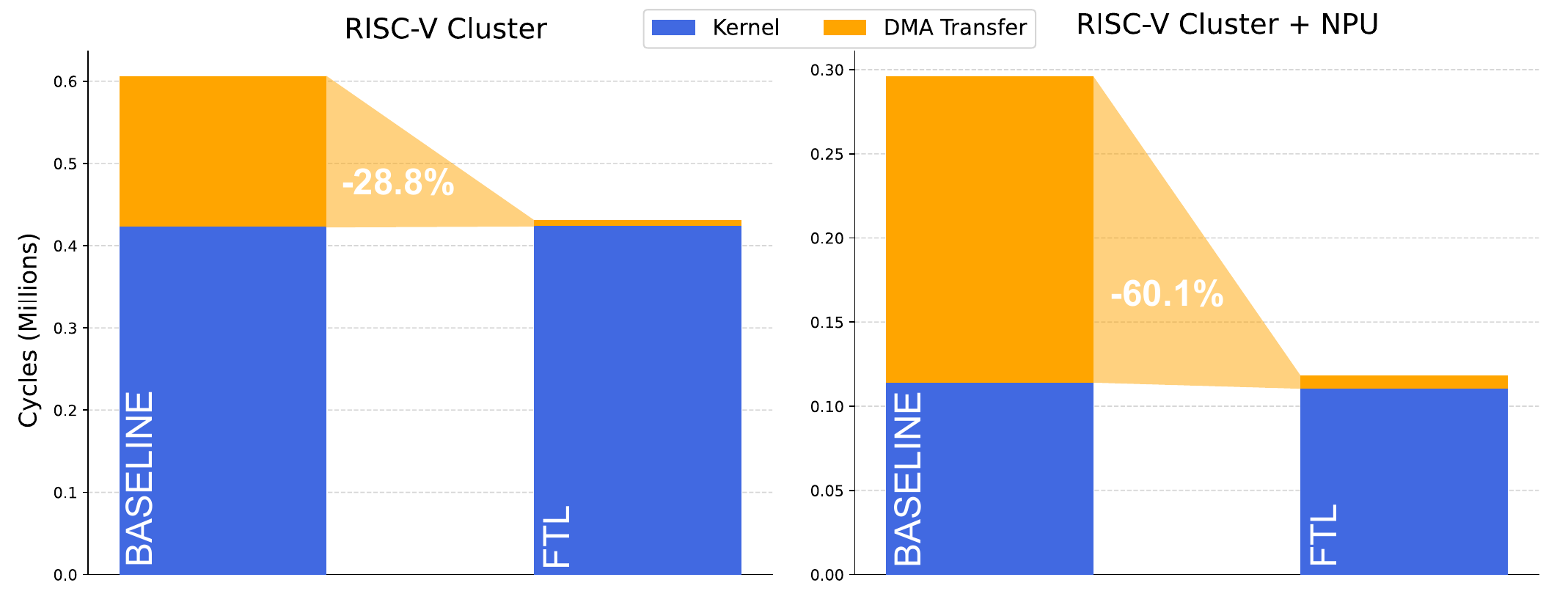}
  \caption{Runtime comparison of \gls{vit}'s \gls{mlp} using layer-per-layer tiling (baseline) and \gls{ftl} on the Siracusa \gls{soc}.}
  \label{fig:benchmark-barplot}
  \vspace{-1.5em}
\end{figure}

To showcase the benefits of \gls{ftl}, we benchmark a \gls{gemm} followed by a GeLU activation function. These layers are commonly found in the \gls{mlp} stage of \glspl{vit}~\cite{dosovitskiy2020vit}.
%
Fig~\ref{fig:benchmark-barplot} reports the \gls{mlp}'s runtime with and without \gls{ftl} when using only the RISC-V cluster (left side) and using the cluster and the \gls{npu} (right side).
There are two reasons to explain such runtime reduction when using \gls{ftl}. First, \gls{ftl} reduces the number of \gls{dma} transfers by 47.1\,\% by preventing the materialization of the \gls{mlp}'s intermediate tensor. Second, the L2 memory capacity is exceeded when materializing the \gls{mlp}'s intermediate tensor; hence, this tensor is stored in L3 \gls{ram}. With \gls{ftl}, we don't need to perform costly off-chip memory transfers to bring back the intermediate tensor from L3 to L1, leading to a reduction of the runtime.
If double-buffering is used, \gls{ftl} speeds up execution only if the kernel runtime is less than the \gls{dma}'s runtime. As reported in Fig~\ref{fig:benchmark-barplot}, this is the case when using the cluster and the \gls{npu}.


\vspace{-1em}
{
\footnotesize
\bibliographystyle{IEEEtran}
\bibliography{bibliography}

\begin{thebibliography}{1}
\providecommand{\url}[1]{#1}
\csname url@samestyle\endcsname
\providecommand{\newblock}{\relax}
\providecommand{\bibinfo}[2]{#2}
\providecommand{\BIBentrySTDinterwordspacing}{\spaceskip=0pt\relax}
\providecommand{\BIBentryALTinterwordstretchfactor}{4}
\providecommand{\BIBentryALTinterwordspacing}{\spaceskip=\fontdimen2\font plus
\BIBentryALTinterwordstretchfactor\fontdimen3\font minus \fontdimen4\font\relax}
\providecommand{\BIBforeignlanguage}[2]{{%
\expandafter\ifx\csname l@#1\endcsname\relax
\typeout{** WARNING: IEEEtran.bst: No hyphenation pattern has been}%
\typeout{** loaded for the language `#1'. Using the pattern for}%
\typeout{** the default language instead.}%
\else
\language=\csname l@#1\endcsname
\fi
#2}}
\providecommand{\BIBdecl}{\relax}
\BIBdecl

\bibitem{chen2018tvm}
{T. Chen et al.}, ``Tvm: an automated end-to-end optimizing compiler for deep learning,'' in \emph{USENIX OSDI`18}.

\bibitem{prasad2024siracusa}
{A. S. Prasad et al.}, ``Siracusa: A 16 nm heterogenous risc-v soc for extended reality with at-mram neural engine,'' \emph{IEEE JSSC}.

\bibitem{dosovitskiy2020vit}
{A. Dosovitskiy et al.}, ``An image is worth 16x16 words: {{Transformers}} for image recognition at scale,'' in \emph{{{ICLR}} 2021}.

\end{thebibliography}
}

\end{document}